\documentclass[preprint,10pt]{elsarticle}

\usepackage{amssymb,amsmath}

\def\be{\begin{equation}}
\def\ee{\end{equation}}

\usepackage{graphicx,eucal}
\usepackage[all]{xy}
\xyoption{all}

\usepackage{color}

\journal{Physica A}

\begin{document}

\begin{frontmatter}



\title{A q-spin Potts model of markets: Gain--loss asymmetry in stock indices as an emergent phenomenon}
\author{Stefan Bornholdt}
\ead{bornholdt@itp.uni-bremen.de}
\address{Institut f\"ur Theoretische Physik, Universit\"at Bremen, Germany}

\begin{abstract}
Spin models of markets inspired by physics models of magnetism, as the 
Ising model, allow for the study of the collective dynamics of interacting 
agents in a market. The number of possible states has been mostly limited 
to two (buy or sell) or three options. However, herding effects of competing 
stocks and the collective dynamics of a whole market may escape our reach 
in the simplest models. Here I study a $q$-spin Potts model version of a 
simple Ising market model to represent the dynamics of a stock market index 
in a spin model. As a result, a self-organized gain--loss asymmetry in the 
time series of an index variable composed of stocks in this market is observed. 
\end{abstract}
 
\end{frontmatter}

\section{Introduction}
This article is dedicated to the memory of Dietrich Stauffer who taught me 
the beauty of simple computer models. He was very supportive of the formation
of a socio- and econophysics community in the german physical society and 
inspired many with his quest for universality in human nature. In this brief 
paper I will study a variation of a spin model for markets which, 20 years 
ago, appeared in the International Journal of Modern Physics C, after 
Dietrich edited and reviewed the manuscript within hours of submission. 
 
Collective effects of traders in a market are of particular interest for the 
dynamics of markets and are notoriously hard to grasp with traditional 
economic equilibrium models of representative agents \cite{Kirman}. Agent 
based models filled the gap in representing the non-rational and collective 
elements of the dynamics of markets \cite{Stauffer,StaufferReview,Kutner2019}. 
Stylized facts of market dynamics, for example, were first reproduced in 
agent based models \cite{LuxMarchesi}. Extreme simplifications of such models,
as in the Ising model variant we come back to here \cite{Bornholdt2001}, 
then allowed to isolate the particular underlying herding mechanism among 
agents, at the roots of stylized facts.  

A curious property of real stock markets is an observed gain--loss asymmetry 
in the time series data of stock indices, but not in the time series of 
single stocks \cite{GainLoss,GainLoss2}. This phenomenon is largely still not
fully understood and could be an interesting case for simple econophysics 
models to test possible mechanisms at the roots of such a collective effect. 
 
We do have an intuitive understanding of why such an observation could be 
plausible: Downward movements of stock markets are often in synchrony, due to
external shocks, or bad news affecting the whole of the market, while, on the
other side, upward movements of stocks happen in a more uncorrelated fashion,
with hopes and fantasy of investors often tied to the ideas and fate of 
single enterprises. A phenomenological model with an external ``fear factor''
that randomly lowers the stocks of a model market from time to time, with 
stocks otherwise performing independent random walks with an upward bias, 
demonstrated that this intuition works \cite{FearFactorModel}. 
 
As a next step, it would be interesting to see how such a phenomenon could 
emerge in a dynamical market model itself. A first model that exhibits the 
gain--loss asymmetry as an emergent effect of its intrinsic dynamics is the 
block-spring conveyor belt model by Bulcs\'u and N\'eda \cite{BlockSpring}. 
It is inspired by the classical self-organized critical block spring model 
originally discussed for earthquakes \cite{CarlsonLanger}. 
 
We here go a step further and ask if this effect could emerge in a 
minimalistic market model inspired by the Ising model of statistical physics
\cite{Bornholdt2001}. However, in a two state spin model a Dow Jones like 
index variable cannot be defined as there is only one stock in this minimal 
version. Let us therefore generalize the spin model market to a q-spin Potts 
model version and define a stock market index variable to study its time 
series. Let us now briefly recapitulate the Ising spin model, then define the
Potts model extension, followed by some numerical experiments. 

\section{A spin market model} 
The attractive nearest neighbor interaction in the ferromagnetic phase of the 
Ising model can be viewed as a simple model for imitating the actions of 
others, a tendency assumed to be at work in the stock market. Adding a second
interaction that represents the opposing tendency of the trader to escape the
crowd, makes a particularly simple model for markets \cite{Bornholdt2001}. 
The magnetic alignment between neighboring spins is countered with an 
additional coupling $\alpha>0$ between the absolute value of the 
magnetization, acting as a destabilizing ''external'' field, to the local 
field of each spin as defined in a local field at spin $S_i$ given by 
\begin{equation} 
h_i = \sum_{j=1}^NJ_{ij} \; S_j - \alpha \; S_i \; \left| \frac{1}{N} 
      \sum_{j=1}^{N} S_j \right|. 
\end{equation}  
The overall model consists of $i=1,...,N$ spins with orientations 
$S_i(t) = \pm 1$ and we consider a 2 dimensional lattice with couplings 
$J_{ij}=1$ for the four nearest neighbors of each spin and $0$ else. The 
dynamics of each spin $S_i(t)$ depends on its local field $h_i(t)$ and is 
determined by a heat-bath dynamics according to  
\begin{eqnarray} 
S_i(t+1) &=& +1 
\;\;\; \mbox{with} \;\;\; 
p = 1/\left[1+\exp\left(-2\beta h_i(t)\right)\right]
\nonumber \\ 
S_i(t+1) &=& -1 
\;\;\; \mbox{with} \;\;\; 
1-p. 
\label{heatbath}
\end{eqnarray}  
The parameter $\beta=\frac{1}{T}$ is the inverse of a formal temperature $T$ 
and determines the heat bath dynamics. 
The newly introduced global coupling constant $\alpha>0$ can be interpreted 
as a "fear" factor, rendering each spin's actions more random, the more 
asymmetry has accumulated in the model market. Note that this term always is 
proportional to $- S_i(t)$, thus pulls the spin to its opposite orientation, 
in particular when magnetization is large. Magnetization can be thought of as 
a convenient indicator of the overall asymmetry, or a ``bubble'' away from 
equilibrium, in the spin orientations. 

When running this model numerically below the critical temperature of the 
Ising model, it does not settle down to an equilibrium state, but rather 
exhibits broad fluctuations with intermittent phases of ``bubbles'' and 
``crashes'', quite similar to the statistical features of financial markets 
called stylized facts. The features also occur in higher dimensions 
\cite{MarketDimensions} and in non-lattice topology versions of the model 
\cite{MarketNetworks1,MarketNetworks2,MarketNetworks3}. A detailed analysis 
of its dynamics has been done e.g.\ in Refs.\ 
\cite{KrauseBornholdt2012,KrauseBornholdt2013}, 
and its multifractal behavior is characterized in Ref.\ 
\cite{Multifractality,Multifractality2}.
Further applications to financial markets and models include
\cite{Horvath2006,Sieczka2008,Meudt2016,Takaishi2016,Takaishi2017,
Kristoufek2018,Olesen2020}.
Discussions in the context of other market models can be found in Refs.\ 
\cite{MarketReview1,MarketReview2,Kutner2019}. 

Some model variants include the introduction of more than binary variables, 
as for example three state models to model the three options of buy, sell, or
hold a stock \cite{Threestatemodel1,Threestatemodel2,Threestatemodel3}. In 
order to model a market with a set of many stocks, an ensemble of spin models 
has been considered \cite{Takaishi2015,Takaishi2017}. However, curiously, a 
version proposed to model markets in the original article 
\cite{Bornholdt2001},the q-spin Potts model, did not catch interest so far. 
Thus let us try this now. 

\section{A q-state Potts model for markets} 
Consider a model with $i=1,...,N$ spins with now $q$ possible states 
$\sigma_i(t) \in \{0,..., q-1 \}$. 
The different states now symbolize not only two but $q>2$ different stocks or 
assets in the model market and each agent will choose one of them to invest 
in (a model where agents can diversify their investments has been explored by
Takaishi \cite{Takaishi2015,Takaishi2017} who considers an ensemble of binary
spin models instead). Let us here consider this minimalistic idealization of 
a market as a $q$-spin Potts model, as the simplest extension of a  buy-sell
spin model to more that two states we can think of. Note that it solely 
focuses on the imitation dynamics of agents, as also the binary spin models 
do, and does not itself contain economic quantities as stock prices. Rather,
a central observable is the popularity of a state which is taken as an 
analogy for the over- or undervaluation of the corresponding asset. Prices 
can be introduced through a market maker mechanism, for example, as has been 
demonstrated for the two state model in \cite{Kaizoji2002}. 

As in the Ising spin model, two neighbors exhibiting the same preference 
(state) will be energetically favored (this is the herding interaction: 
``buy the stock that your friends buy''), and this is already conveniently 
expressed in the q-state version of the original Potts model 
\cite{Potts:1952,RevModPhys.54.235} as defined by the Hamiltonian 
\begin{eqnarray} 
	H &=& - J \sum_{<ij>} \delta_{\sigma_i\, \sigma_j} 
	\\ \nonumber 
	\;\;\; \mbox{with} \;\;\; 
	\delta_{\sigma_i\, \sigma_j} &=& 1 
	\;\;\; \mbox{if} \;\;\; 
    \sigma_i = \sigma_j
    \\ \nonumber 
    \;\;\; \mbox{and} \;\;\; 
    \delta_{\sigma_i\, \sigma_j} &=& 0
    	\;\;\; \mbox{if} \;\;\; 
    \sigma_i \not= \sigma_j.
    \label{potts}
\end{eqnarray}  
Let us always set $J=1$. 
For $q=2$, this model maps onto the Ising model with $S_i = 2\sigma_i(t)-1 $ 
and $\sigma_i(t) \in \{0,1 \}$. To see this, first rewrite the heatbath 
update (\ref{heatbath}) as a flip probability 
\begin{eqnarray} 
p_{flip} (S_i \rightarrow - S_i) 
&=& \frac{1}{1+\exp\left(2\beta S_i(t) h_i(t)\right)}.
\label{isingflip}
\end{eqnarray}  
For the Ising spins $S_i(t)$, this corresponds to the heatbath flip 
probability
\begin{eqnarray} 
p_{flip} &=& \frac{1}{1+\exp\left(\beta \Delta E\right)} 
\end{eqnarray}  
as a function of the energy difference 
\begin{eqnarray} 
\Delta E =  2 S_i(t) h_i(t). 
\end{eqnarray}  

In the Potts model, for each single spin update we choose a site $\sigma_i$ 
and randomly pick a state $\sigma_i(new)= \mu$ with $\mu \in {0,...,q-1}$. 
The new local energy minus the old local energy of this site with respect to 
its 4 neighbors then is 
\begin{eqnarray} 
\Delta E &=& - \sum_{j\in nn(i)} \left( \delta_{\sigma_i(new)\, \sigma_j} 
             - \sum_{j\in nn(i)} \delta_{\sigma_i(t)\, \sigma_j}\right).  
\label{pottsglauber}
\end{eqnarray} 
The flip probability in the q=2 Potts model formulation then is 
\begin{eqnarray} 
p_{flip} (\sigma_i(new)) &=& \frac{1}{1+\exp\left(-2\beta \left[
2\sum_{j\in nn(i)} \delta_{\sigma_i(new)\sigma_j(t)} - 4 \right]\right)}. 
\label{flipprob4model} 
\end{eqnarray}  
With $S_i = 2\sigma_i(t)-1 $ this can be rewritten as 
\begin{eqnarray} 
p_{flip} (S_i(new)) &=& \frac{1}{1+\exp\left(-2\beta S_i(new) 
                                             \sum_{j\in nn(i)} S_j\right)}. 
\end{eqnarray}  

This is equivalent to (\ref{isingflip}) thus, for $q=2$, the Potts and Ising 
models yield the identical local field and heatbath updates when both 
formulated with spins.  
\\

Unfortunately, this equivalence does not hold for the $q>2$-state Potts model, 
so let us briefly recapitulate what this means for the local interaction 
``buy the stocks that your friends buy'' in our market model. 

The Ising heatbath flip probability can be read in two ways: 
\begin{enumerate}
\item The flip probability is determined by the energy difference $\Delta E$ 
of the flip.
\item The flip probability is determined by the new local energy 
$\sum_{j\in nn(i)} \delta_{\sigma_i(t+1)\, \sigma_j}$ after the flip.
\end{enumerate}
This is equivalent for $q=2$, as we have 
$\Delta E = 2 \sum_{j\in nn(i)} \delta_{\sigma_i(t+1)\, \sigma_j}$. 
Note that for 
$q>2$, $\Delta E \not= 2 \sum_{j\in nn(i)} \delta_{\sigma_i(t+1)\, \sigma_j}$,
\\
but instead $\Delta E = \sum_{j\in nn(i)} \left[
\delta_{\sigma_i(t+1)\, \sigma_j} - \delta_{\sigma_i(t)\, \sigma_j} \right]$. 
From a physical point of view, the straightforward way to generalizing the 
model to $q>2$ is to use the Glauber update \cite{Glauber} probability for 
any general move that changes the energy of the system by some $\Delta E$    
\begin{eqnarray} 
p_{flip} &=& \frac{1}{1+\exp\left(\beta \Delta E\right)}. 
\end{eqnarray}  

However, is this interpretation as a local energy budget an accurate account 
of what happens in the market context, and of what a trader does when 
flipping one stock to another? One could argue that it reflects the 
accounting of selling one stock and buying another. Yet, when modeling market
dynamics of traders without explicit fundamental knowledge of the stocks, but
speculating on future expectations instead, standard accounting is most 
surely the wrong perspective, as also the more traditional financial market 
models indicate in failing to reproduce stylized facts in their dynamics. So 
what could be the underlying microscopic interaction that leads to the 
formation of expectation bubbles? 

New opportunities are best represented by the future characteristics of what 
we consider opting for, therefore let us choose the second interpretation of
the above options when extending our model market to more than two stocks 
options: This rule favors the new over the old, in fact even when the energy
change associated to the flip is zero. 

In practical terms, for $q>2$, we will apply the $q=2$ heatbath update rule 
of our above Potts version (\ref{flipprob4model}) of the Ising model, solely 
being a function of twice the energy contribution from the new spin 
orientation. Note that this is not the energetically accurate Glauber flip 
probability, as the old energy is not subtracted. However, throwing energy 
conservation overboard, we gain the freedom to incorporate the pull of the 
future with this step. 

To construct a market model, it remains to complement this herding 
interaction by a counteracting term that represents the fear of the trader in
the face of a market bubble. In the Ising model version, a large 
magnetization is the indication of such a disequilibrium in the market.

The standard order parameter of the $q$-spin Potts model \cite{Binder1981} is
based on the number of spins $N_{max}(t)$ exhibiting the state that occurs 
most frequently at a given time $t$ of the simulation
\begin{equation}
M(t) = \frac{q\frac{N_{max}(t)}{N}-1}{q-1}.
\end{equation}
From the perspective of a stock market, focusing attention on the largest 
stock may be not completely unrealistic. In our model, let us  therefore 
choose this Potts model order parameter for the second term in the local 
field, coupling the agent's fear of a bubble to the stock with the largest 
value (or perhaps the largest overvaluation, when thinking of bubbles) in the
market. 

Thus the full $q$-spin market model is given by  
\begin{eqnarray} 
h_i &=& 2\sum_{j\in nn(i)} \delta_{\sigma_i(new)\sigma_j(t)} - 4 - \alpha M(t).
\\
M(t) &=& \frac{q\frac{N_{max}(t)}{N}-1}{q-1}
\\
p_{flip}(\sigma_i(new)) &=& \frac{1}{1+\exp\left(-2\beta h_i\right)}. 
\end{eqnarray}  
After random initialization of the lattice, spins are serially updated in 
random order, time is given in number of update steps per spin (sweeps).  
Temperature has to be chosen below the critical temperature. Note that while 
the critical point of the Potts model as defined in (\ref{potts}) is given by
$T_c = 1 / \log(1+1/ \sqrt{q})$ ($J=1$), its $q=2$ adaptation to match the 
Ising model done in (\ref{flipprob4model}), however, has the extra factor of 
2 of the critical temperature of the Ising model $T_c=2/ \log(1+1/ \sqrt{2})$. 
As we use (\ref{flipprob4model}) for our model even with $q>2$, it is an 
interesting question how this generalizes to $q>2$. Numerically one observes 
a phase transition different from the standard Potts model transition and at
a larger temperature than the critical temperature in the Potts model.  
For the market model, we need to operate well below the critical temperature. 
Also when switching to the non-physical update rule as compared to the 
original Potts model, we are on the safe side as that variant is still below 
critical even at the critical point of the original Potts model, as can be 
easily seen by switching to the physical Glauber update rule 
(\ref{pottsglauber}). For our simulations we choose $T=0.2 T_c$ of the Potts 
model. 
\begin{figure}[htb]
\centering
\includegraphics[scale=0.2]{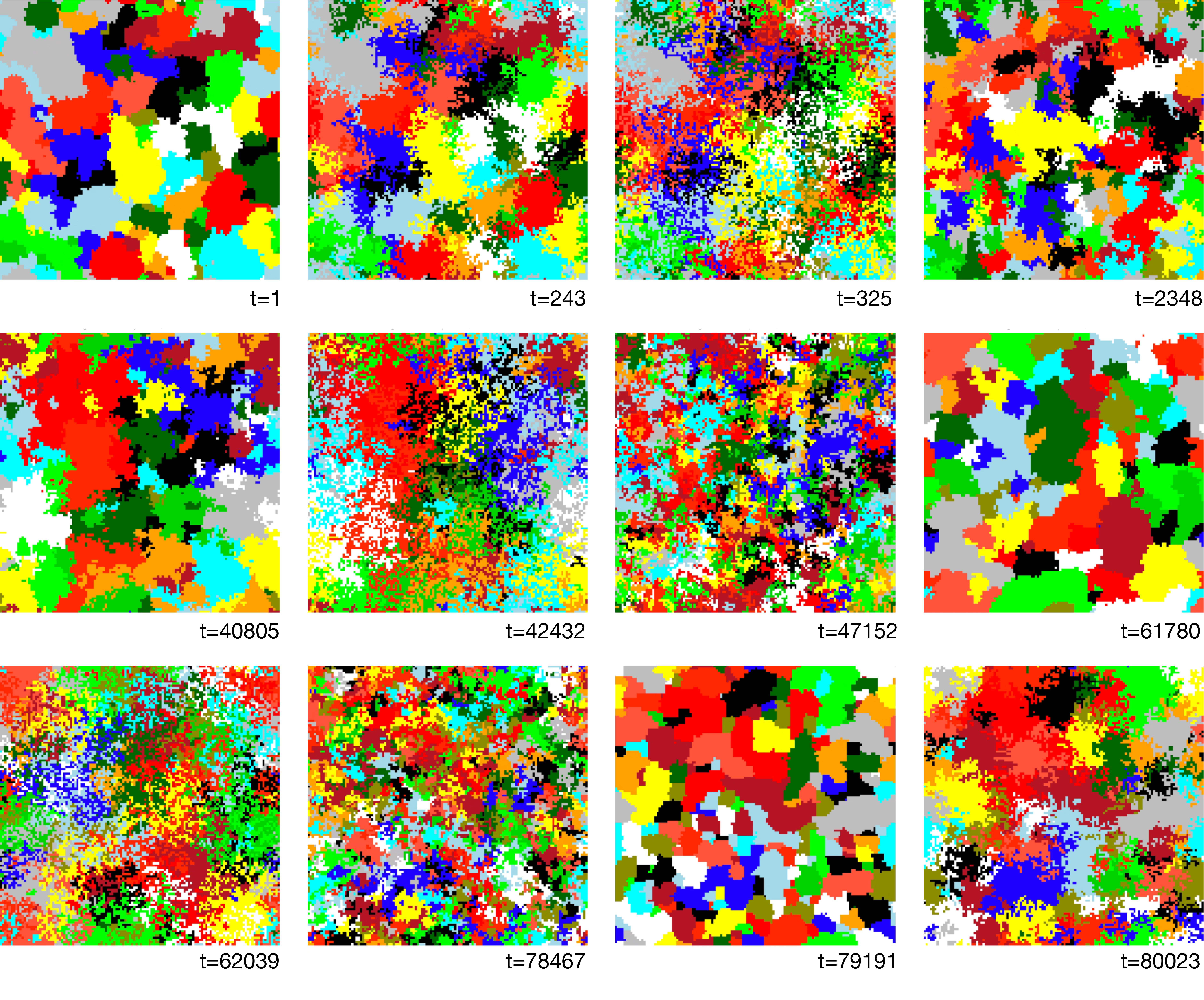}
\caption{The $q$-state Potts model inspired market model. 
Simulation on a 128x128 lattice with $q=16$ possible states, 
temperature $T = 0.33$, and ``fear''-parameter $\alpha = 100$.}
\label{pottsmarket}
\end{figure}
Figure \ref{pottsmarket} shows a simulation of this model under random serial
single spin update. 

\section{Dynamics of the model} 
The dynamics has similarities with the two state Ising market model, namely 
an intermittent behavior of calm and chaotic phases. 

Note that the dynamics of the model at a temperature well below the critical 
temperature is quite different from the original Potts model which, below 
$T_c$, settles with one of the $q$ states as the dominant state. Instead our 
dynamical ``rest'' state is one where each of the $q$ states occupies 
approximately $N/q$ states of the system, typically occurring in patches, 
each dominated by a single state. Fluctuations away from this state are 
penalized with a higher flip probability (or effective temperature {\em sensu}
\cite{KrauseBornholdt2013}) through the second term in the local field of 
each agent. Deviations from the symmetric default state exhibit similar 
activity avalanches as we see in the Ising spin market model. 

\begin{figure}[htb]
\centering
\includegraphics[scale=0.5]{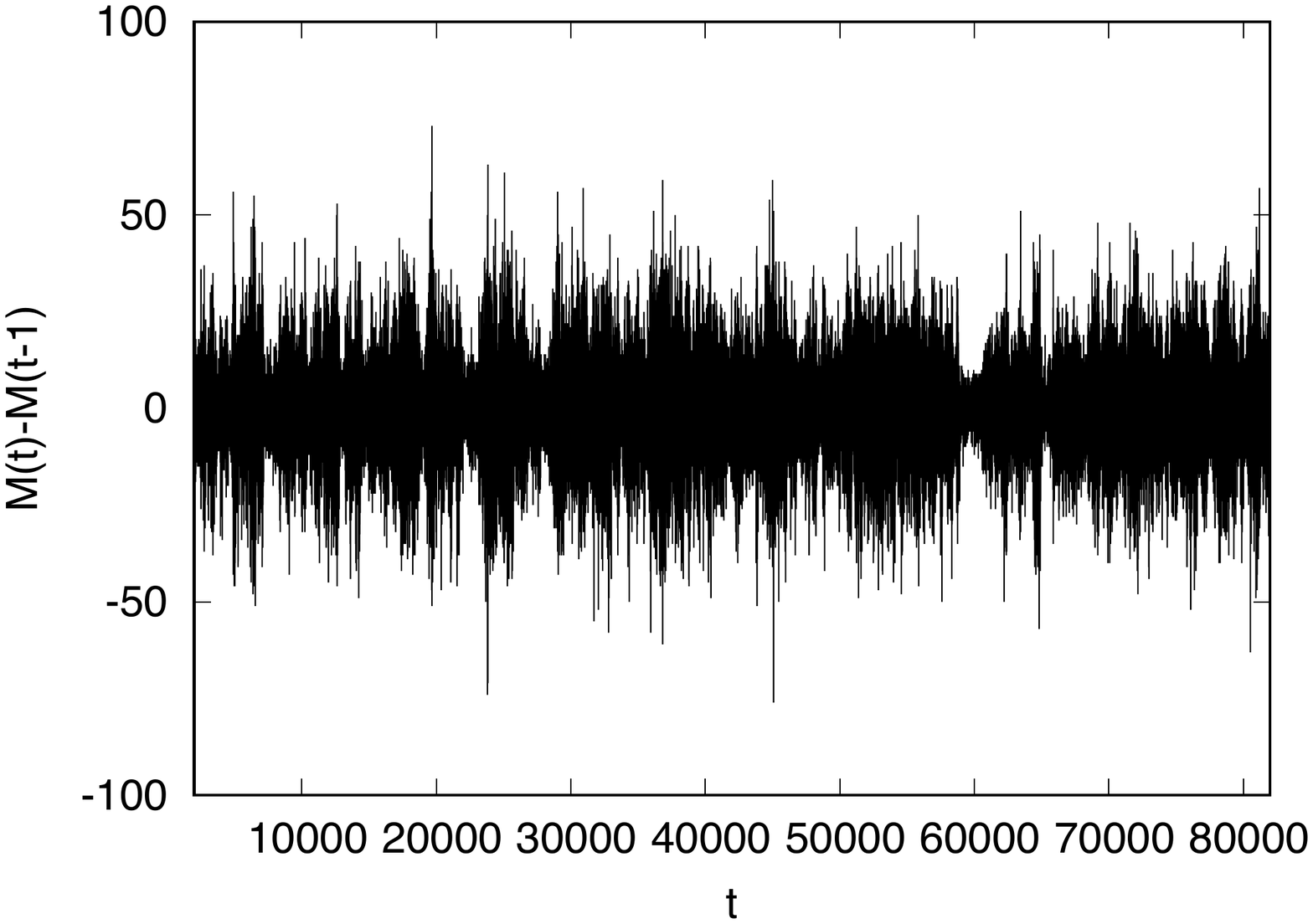}
\caption{The change per time step of the order parameter $M(t)$ in the 
$q$-state market model. Simulation on a 128x128 lattice with $q=16$ possible 
states, temperature $T = 0.33$, and fear parameter $\alpha = 100$.}
\label{orderparameter}
\end{figure}

An overall indicator is the order parameter $M(t)$, see the changes of it per
time step in figure \ref{orderparameter}.
One observes intermittent phases with higher and lower volatility, quite 
similar to, yet not as pronounced as in the Ising market model (which is also
the $q=2$ case of the current model). While the latter \cite{Bornholdt2001} 
reproduces some of the prominent statistical characteristics of financial 
markets ("stylized facts") quite well \cite{Multifractality}, the $q=16$ 
example of the new model does not. As this article appears in a volume to the
memory of Dietrich Stauffer, let us remember that Dietrich always insisted
that also negative results should be published. 

Therefore, let us consider this an interesting observation that deserves 
further discussion. It may find a technical explanation from the perspective 
of our analysis of the spin market model \cite{KrauseBornholdt2013}. There we
found, for $q=2$, dominant subcritical metastable striped states in the 
systems dynamics with a vastly fluctuating borderline length that separates 
the two states (occurring as adjacent stripes) on the lattice, with very 
large interface lengths in the fragmented phases. For $q>2$, we do not have 
the situation of the striped states, instead, the calm phases are already 
exhibiting a much larger interface borderline length around the small patches
of different $q$-values. In the binary case, a small temperature $T$ drives 
the system down to the striped state while, at the same time, a large $\alpha$
enforces strong intermittency. For large $q$, the equivalent of striped 
states can be observed for a small $\alpha$, where a majority of the $q$ 
states disappear, but then $\alpha$ cannot at the same time be made large to 
obtain strong intermittency. 

With these remarks we leave the question of stylized facts in the new model 
class and possible modifications for future research and now focus on 
collective effects in the dynamics of a toy model stock index, which can be 
defined in the Potts stock model. 

\section{The dynamics of a market index} 
\begin{figure}[htb]
\centering
\includegraphics[scale=0.6]{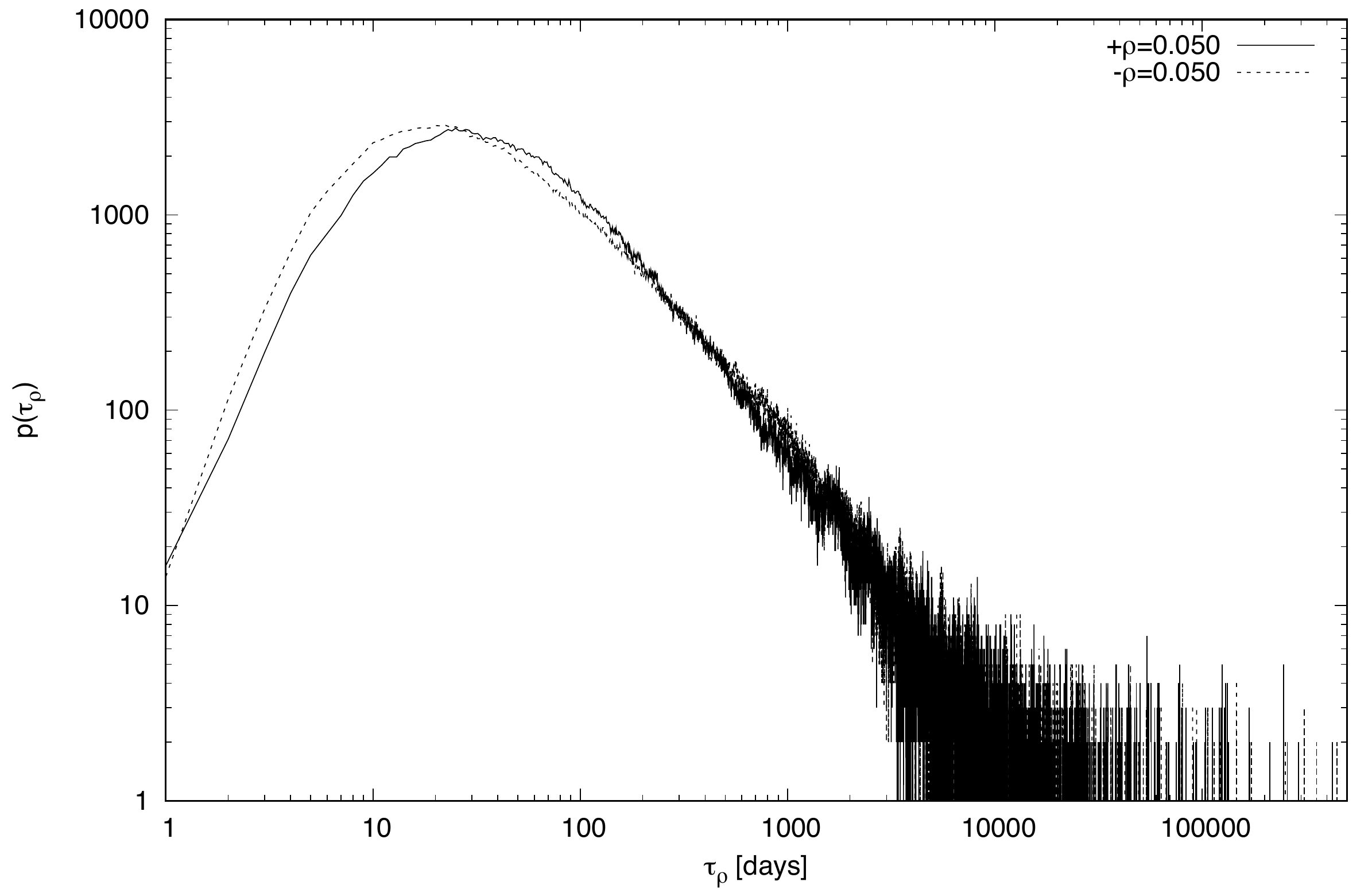}
\caption{The probability distributions $p(\tau_\rho)$ of waiting times 
$\tau_\rho$ for positive and negative relative changes of $\rho = \pm 0.05$
as derived from the index time series $I(t)$. 
Note that the loss occurs on average earlier than the gain of same size. 
Simulation parameters are the same as in Figure 1.}
\label{index}
\end{figure}
Let us finally define a market index for our model market. With $q$ 
different competing states in the system which fluctuate around a default 
value, we consider these as different stocks in a market fluctuating around
their fundamental value. An index of a financial market is commonly a 
selection of stocks. Unsuccessful ones drop out of the index and are 
replaced by new ones with potential to grow. To mimic this in the simplest 
way, we define our model index to comprise those of the $q$ stocks, that 
have a higher than average share. We include stock $\mu$ once its share is 
larger than $N_\mu(t) > N/q$, thus define our index by 
\begin{equation}
I(t) = \sum_{\mu=0}^{q-1} (N_\mu(t)-N/q) \; \Theta(N_\mu(t)-N/q)
\end{equation} 
with the Heaviside function $\Theta(x)=1$ for positive argument $x$, and 
else $\Theta(x)=0$. While real stock indices usually consist of a fixed 
number of stocks, we here choose the threshold definition solely for 
computational simplicity. Note that it does not affect the dynamics of the 
model, as it is just a suitably defined observable. 
 
With this index we can now analyze the gain--loss asymmetry of the time 
series. Figure \ref{index} shows the distribution of waiting times for a 
five percent gain or loss in the index. Both waiting time distributions 
have a maximum at a certain number of trading days, however, the most 
likely waiting time for a five percent loss occurs earlier (at about 20 
trading days) than the most probable waiting time for a five percent gain 
(about 30 trading days). This is the main observation of this study. 

Another feature of these distributions is a power law decay towards larger 
times, as one expects for first passage waiting time distributions of 
random walks. Note that for large waiting times $\tau_\rho$, the asymptotic
probability distribution  $p(\tau_\rho) \propto \tau_\rho^{-\theta}$ 
follows an exponent of approximately $\theta = 4/3$ which is a lower value
than the value $\theta = 3/2$ for a random walk. The lower value of $\theta$
means that the probability of longer waiting time intervals is increased, 
pointing to a superdiffusive movement of the index variable $I(t)$. 

\section{Discussion}
The modified $q$-state Potts model we use here as a toy model for markets 
exhibits a gain--loss asymmetry as is well known from real stock indices. 
The interpretation is that the model generates a synchronized downward 
movement across the range of stocks in the index, through the instability 
inherent to the model. This is a self-organized version of the Donangelo et
al.\ model \cite{FearFactorModel} which used an external ``fear factor'' 
instead. 

This first small study of a $q$-state Potts model for a market leaves many 
questions open. The simplistic view of returns in terms of the changes in 
relative frequencies of $q$-state occurrences among agents has to be 
further interpreted in a market context for a full understanding. A price 
definition, for example via a market maker \cite{Kaizoji2002}, would be a 
first step. 

Another open question is if and how stylized facts of financial markets 
could be reproduced in a model with more than $q>2$ states. Earlier 
analyses of the Ising spin market model could be a starting point, as 
outlined above. Further development of the model may also involve the 
definition of the local field $h_i$ in equation (12) and how it depends on 
the order parameter $M(t)$. I found it crucial for the dynamics that the 
nature of the new state $\sigma_i(new)$ does not depend on the majority 
Potts state. However, this might contradict the intuition of how traders 
act in real markets and other definitions of $h_i$ could turn out to be 
interesting. 

We here studied the model on a two dimensional lattice, mainly because the 
binary Ising-model version of the market model has been thoroughly analyzed
in the two dimensional version and forms an interesting background model, 
where interface length plays a central role in the model. Alternatively, 
the regular lattice could be replaced with a (random) graph, opening up for 
questioning the role of topology and even for the possibility of evolving 
topologies. 
 
Last, not least, different socio-economic phenomena or systems, other than
markets, could be addressed with this approach as, for example, Potts model 
versions of voter models for dynamics of opinion formation.

\section*{Acknowledgments}
I thank H.\ Beushausen for valuable discussions. 

\bibliographystyle{plain}

\end{document}